\documentstyle[12pt,aasms4]{article}

\begin{document}

\title{The Absolute Proper Motion of Palomar 12: A Case for Tidal 
Capture from the Sagittarius Dwarf Spheroidal Galaxy}
\author{Dana I. Dinescu\altaffilmark{1}, Steven R. Majewski\altaffilmark{1,4}, Terrence M. Girard\altaffilmark{2} and 
Kyle M. Cudworth\altaffilmark{3}}

\altaffiltext{1}{Department of Astronomy, University of Virginia, Charlottesville, VA 22903-0818 (dd7a@virginia.edu, srm4n@virginia.edu)}
\altaffiltext{2}{Astronomy Department, Yale University, P.O. Box 208101,
New Haven, CT 06520-8101 (girard@astro.yale.edu)}
\altaffiltext{3}{Yerkes Observatory, P.O. Box 258
Williams Bay, WI 53191-0258 (kmc@yerkes.uchicago.edu)}
\altaffiltext{4}{Visiting Associate, Observatories of the Carnegie Institution of Washington, David and Lucile Packard Foundation Fellow, Cottrell Scholar}

\begin{abstract}
We have measured the absolute proper motion of the young globular cluster
Pal 12 with respect to background galaxies, using plate material
spanning a 40-year time baseline, and measuring stars
down to a magnitude $V\sim 22$. The measured 
absolute proper motion has an uncertainty of $0.3$ mas yr$^{-1}$ in each
coordinate.

Pal 12's young age for a globular cluster led to the hypothesis that the
cluster originated in the Large Magellanic Cloud (LMC) and  was later captured by the Milky Way (Lin \& Richer 1992).
Here we investigate this hypothesis using the complete kinematical data.
We present the orbital characteristics of Pal 12 and compare them with
those of the LMC and Sagittarius dwarf galaxy (Sgr). 
The present kinematical data suggest that, from the two parent
candidates for Pal 12, Sgr presents a more plausible case for the host galaxy than the LMC.

We explore this scenario in the context of the 
uncertainties in the orbits and using two different analyses:
the direct comparison of the orbits of Pal 12 and Sgr as
a function of time, and 
the analytical model of Sgr's tidal disruption developed by
Johnston (1998).
We find that, within the present uncertainties of the
observables, this scenario is viable in  both methods. Moreover,
both methods place this event at the same point in time.
Our best estimate of the time of Pal 12's tidal 
capture from Sgr is $\sim 1.7$ Gyr ago.

\end{abstract}


\section{Introduction}

Palomar 12 (Pal 12,
$\alpha_{2000} = 21^{h}~46\farcm6,~\delta_{2000} = -21\arcdeg~
15\arcmin$, l = $30\fdg5$, b = $-47\fdg7$) is now well known as a
young globular cluster with a tidal
radius of $7\farcm6$ and a concentration parameter of 1.08
(Rosenberg {\it et al.} 1998).

Extensive studies of the color-magnitude diagram (CMD)
starting with 
Harris \& Canterna (1980) and continuing with Gratton \& Ortolani
(1988), Stetson {\it et al.} (1989, hereafter S89) and 
Rosenberg {\it et al.} (1998, hereafter R98) indicate that Pal 12
is between $25\%$ and $68\%$ younger than the majority of the globular
clusters of our Galaxy.
Moreover, recently, Brown {\it et al.} (1999) have measured the ratio of
$\alpha$-processed elements  to iron, and they
found that [$\alpha$/Fe] = 0.0 for Pal 12, whereas the bulk of halo
globulars have [$\alpha$/Fe] = 0.3. Thus Pal 12 must have formed in an
environment where type Ia supernovae dominated, which agrees with its 
young age based on the CMD.
In the context of the Searle \& Zinn (1978) 
Galaxy-formation picture, such young clusters, found mostly  in the
outer halo, were formed in the fragments or satellites
that evolved independently from the main body of the Galaxy
(see also Zinn 1993).
These considerations together with the large spread
in the ages of globular-type clusters in the 
LMC (Elson \& Fall 1988, Sarajedini 1998), 
led Lin \& Richer (1992, hereafter LR92) to explore the possibility 
that Pal 12 was captured
from the LMC, during a pericenter passage of the Cloud. 
The LR92 study, which was based on location and radial-velocity data,
made a tentative prediction for Pal 12's 
proper motion, but admitted that among the two clusters
studied --- Ruprecht 106 and Pal 12 --- the latter is less likely to 
have been captured from the Cloud.

The purpose of this work is to measure the tangential velocity
of Pal 12 and determine whether the above-mentioned 
scenario is supported by the kinematical data. Our study of Pal 12 continues
a program (Majewski \& Cudworth 1993) to derive proper motions
of a number of distant globular clusters and dwarf spheroidals.

Sections 2, 3 and 4 describe the photographic plate material, the
photometry and the astrometry  respectively. In Section 5 we 
present the correction to absolute proper motion based on
galaxies. In Section 6 we compare
orbital characteristics of Pal 12 with those of the LMC and Sgr.
Section 7 presents a detailed investigation of the Pal 12 and Sgr
orbits as a function of time, a comparison between the 
predicted present positions and velocities for tidal streams from Sgr 
(Johnston 1998, Johnston {\it et al.} 1999, hereafter J99)
and those of Pal 12, and the properties of cluster Pal 12 
in comparison with those of the clusters associated with Sgr.
In Section 8 we summarize the main conclusions of this paper.

\section{Observational Material and Measurements}

In our proper-motion study of Pal 12, we 
have used a collection of 31 photographic plates taken at three
different epochs. These are:
ten plates taken with the Las Campanas-DuPont (LCO) 2.5-m reflector
(scale = $10\farcs92$ mm$^{-1}$, epoch $\sim$ 1993), eleven plates taken with 
the CTIO 4-m reflector (scale = $18\farcs60$ mm$^{-1}$, epoch $\sim$ 1976,
UBK7 corrector),
 and ten plates taken with the Hale 5-m reflector (scale =
 $11\farcs12$ mm$^{-1}$, epoch $\sim$ 1954). 
The characteristics of these plates are summarized in Table 1.
The area covered by our measurements is $\sim 0.5$ deg$^{2}$; however
the most precise proper motions, which include the oldest
plates (Hale 5-m), are limited to an area of 15$\arcmin$ x 15$\arcmin$. 

We have prepared an input catalog
using the deepest, best-quality modern plate (CD3018, see Table 1).
This plate was digitized with the University of Virginia
PDS microdensitometer (30-$\mu$m pixel 
size), and preliminary positions, object diameters and
object classification  were determined using the
FOCAS software (Valdes 1982, 1993). 
Then, all of the plates were measured
with the Yale PDS microdensitometer in a fine-raster, object-by-object
mode, in which 
the input positions and raster sizes were calculated from  the
preliminary catalog obtained from plate CD3018.
We have used a pixel size of
$12.7~\mu$m for the DuPont 2.5-m and Hale 5-m plates, while for the 
CTIO 4-m plates we have used a 10-$\mu$m pixel, as the plate scale is 
slightly larger for these latter plates. The image positions 
on each plate were determined using the Yale Image Centering routines
(two-dimensional, bivariate Gaussian fit, Lee \& van Altena 1983). 
Due to the thermal drift in the PDS during long scans,  
seven stars were repeatedly measured in order to monitor and correct for 
drifts in the measurement system. This correction includes terms for 
translation and rotation.
The typical image-centering accuracy for well-measured, stellar objects
ranges between 0.8 and 1.3 $\mu$m, depending on the plate emulsion.

All of the objects classified
as galaxies by the FOCAS software were visually inspected and an
accurate list of galaxies was redetermined.
Our study is complete down to a magnitude of  $V \sim$ 21.3.

\section{Photometry}

In order to determine our $BV$ photometry we have used the instrumental
magnitudes obtained from the Yale scans of the Las Campanas 2.5-m plates,
and the calibration sequence given from a combination of photoelectric 
and CCD photometry. The photoelectric photometry (Harris \& Canterna 1980)
ensured a good calibration for the bright magnitude range ($V = 11$ to 18); we have 
used 39 stars in common. The CCD photometry included the $BV$ 
photometry from S89, which constrained the
calibration towards the faint end ($V = 14$ to 22). However this study
covered a relatively small area on our plates, and therefore we have also
included only the $V$ magnitudes from the $VI$ CCD photometry of 
R98. This latter study covered a larger 
area and, together with the photoelectric photometry, allowed us to explore the
variation in the photometric calibration across the plate.
An offset of $\Delta V = 0.05$ mags was found 
between  S89 and R98, and we
have applied this offset to all $V$ magnitudes from R98; 
also for stars in common we have preferred the 
S89 photometry as it has a better resolution, and 
it is more appropriate in the crowded region of the cluster.
No offset was applied between the Harris \& Canterna (1980) photoelectric
photometry and S89 CCD photometry, as these offsets 
are negligible (see the comparison in S89). We have 
a total of 457 calibrating stars in $B$ and 875 in $V$.

For each plate we have derived a calibrating curve in the 
appropriate passband. The calibrating curve is determined by cubic spline
interpolation.
We have found that this method provides a more 
appropriate representation of the calibrating curve than the 
traditional fit with one polynomial.
We have also examined the photometric 
residuals as a function of the position on the plate and we have found
a small linear variation for a handful of plates. We have applied this
correction (of the form $a_{x,y}+b_{x,y}x,y$, where $x$ and $y$ are the 
coordinates on the plate) whenever the gradient across 
the plate was significant
(the term is larger than $1.5\sigma$, where $\sigma$ is the 
uncertainty in the term). The largest value of this gradient was 
0.0025 mag mm$^{-1}$, which amounts to a difference of 0.4 mag between the 
edges of our field. Typical random photometric errors per star and plate
are of the order of 0.1 mag at $V \le 19$ and they increase rapidly with
magnitude. Calibrated magnitudes for each plate were averaged to obtain
a final value, and $B-V$ colors were determined as straight differences of
these averaged calibrated magnitudes. We have 
plotted the differences in our final photometry and the standard photometry
($\Delta V$ and $\Delta (B-V)$) versus magnitudes, colors and
positions on the plate, and found no  significant trends.
From the scatter in these differences we conclude that our $V$ magnitudes 
have an uncertainty of 
0.07 mag, and our $B-V$ colors an uncertainty of 
0.1 mag for stars brighter than $V=19$. For magnitudes $V \sim 21$,
uncertainties of 0.2 mag in colors are not uncommon, especially in the cluster
area where crowding affects our photometry.

\section{Astrometry}

All of our three sets of plates (Table 1) were taken with reflectors 
that have significant distortion of various magnitudes. Fortunately
these optical systems have been studied, and empirical calibrations
of the distortion based on astrometric standard fields have been
published. In our investigation we have  used the study of
Cudworth \& Rees (1991) for the Las Campanas 2.5-m and
the CTIO 4-m plates, and Chiu (1976) for the Hale 5-m
plates, to precorrect for distortion.
The Las  Campanas 2.5-m plates were also precorrected for 
differential refraction (third-order refraction theory: Taff 1981), 
before the correction for distortion. The atmospheric refraction 
changes the relative distance between images on the same plate and it is
primarily dependent on the zenith distance. Since only in the case of
the Las Campanas 2.5-m plates the amount of refraction correction is 
comparable to the amount of distortion correction, we have precorrected 
only these plates for refraction. 
Table 2 summarizes the size of the 
correction with respect to the plate center (in arcsec) for
differential refraction (at an hour angle = 0.0), 
and for distortion, for each of the
three telescopes, at two values of the radius from the plate center: 
$5 \arcmin$ and $10 \arcmin$.
The CTIO 4-m and the Hale 5-m 
plates have very strong distortion, and the uncertainty in the 
distortion itself (uncertainties in the coefficients and in the position of the
center of distortion; see Table 3 from Cudworth \& Rees 1991)
is of the size of the refraction correction. 
Since our main purpose is to obtain proper motions free
of systematics, and not necessarily accurate absolute positions, 
we have chosen not to precorrect the positions
for differential refraction
derived from the CTIO 4-m and Hale 5-m plates, but rather let the
leftover geometric systematics be absorbed in the plate model.

The magnitude-dependent systematics in the proper motions (the 
``magnitude equation'')
were treated in the Yale-developed procedure, in which the
cluster stars are used to model this correction, because they 
represent a system with a common motion.
Thus, using the cluster stars, one can separate the guiding-induced
magnitude systematic in proper motions, from the true, secular proper-motion
effect (see details in e.g., 
Guo {\it et al.} 1993, Dinescu {\it et al.} 1996, 
Galadi-Enriquez {\it et al.} 1998).
Since the cluster covers a magnitude range from $V = 14.5$ down to
the plate limit, we can provide a reliable magnitude-equation correction
only in this range, with some uncertainty toward the bright end, where 
there are few cluster stars  to model the systematics.
The preliminary list of cluster stars that defined the magnitude equation,
 was selected via positions in the CMD (S89, R98). 

The relative proper motions were derived using an iterative 
central-plate-overlap algorithm (see for instance Girard {\it et al.} 1989).
The plate model included polynomials of up to fourth order, and linear
color terms. No measurements from the Hale 5-m plates were 
considered in the plate models and final proper motions
if the object resided outside 
a square box of 80 mm on a side (or $14\farcm6$) centered on the
plate center. Due to coma, the stellar images
degrade very rapidly with  distance from the Hale-5m plate center, and 
no reasonable plate model is able to reproduce the distorted positions,
since high-order geometric terms are needed in a region where 
there are very few stars to determine them.
In addition to this strong distortion, the Hale 5-m plates display
very strong color terms. While the majority of the plates 
have color terms of the order of up to a few $\mu$m mag$^{-1}$, the Hale
5-m plates have color terms between a factor of two and a factor of ten
larger than the other plates.
The size of the color terms on the Hale 5-m plates in the $x$ direction, 
which is aligned with the 
right ascension, is correlated with the hour angle.
Therefore these color terms are due to differential color refraction
that is probably significant at the large zenith distance of 
these observations. However, the largest color terms (10 to 15 
$\mu$m mag$^{-1}$) were 
detected in the $y$ direction (aligned with the declination), and they show
no correlation with the hour angle. 
Thus a careful analysis is required if one wishes to include the Hale
5-m plates, which, in spite of their poor astrometric quality,
offer an excellent time baseline and plate scale.

We have reduced all the Las Campanas 2.5-m plates into a common system,
which used plate CD3018 as the standard. This plate was
chosen because it was taken at the lowest hour angle, and
therefore has minimal color terms due to
possible color refraction. It is also one of the deepest, best
image-quality plates. Since, at this point, we are working 
with plates of the same epoch, all stars
used in the plate transformation --- and not only cluster stars ---
can define the magnitude equation. 
However, in order to handle properly
the color systematics in their relation to the magnitude systematics,
we have used only a narrow range in color ($ 0.3 \le (B-V) \le 0.7$)
to define the magnitude equation. This is because, for plates with 
significant color terms,  blue stars may have a different magnitude equation 
than red stars.  Simply defining the magnitude equation based on all stars
may introduce artificial color terms, as --- given the 
morphology of the cluster in the CMD --- we do not have the same distribution in color at all 
magnitudes. Under the assumption that the magnitude equation for 
blue stars differs from that of red stars only by an offset that is
proportional to the color difference, a linear 
color term should subsequently account for the color effects independently
of the magnitude equation. For details of this approach see
Kozhurina-Platais {\it et al.} (1995).
Positions from all 1993 plates were then averaged to form a 
``mean 1993'' plate.
Similarly, we have chosen plate \# 2328 as our
1976 standard for the CTIO 4-m plates, and have formed a ``mean
1976'' plate. Subsequently, the mean 1976 plate was 
transformed into the mean 1993
plate, with the magnitude equation defined, this time,
by cluster stars in the $ 0.3 \le (B-V) \le 0.7$ range.
This transformation provided preliminary proper motions that are
free of magnitude/color systematics, and that were subsequently
used to model all of the plates, including the Hale 5-m plates.

For each star the proper motion is calculated from a linear least squares 
fit of positions as a function of plate epoch. The error in the proper motion
is given from the scatter about this best-fit line. Measurements that
differ by more than 18 $\mu$m ($0\farcs2$) from the  best-fit line
were excluded.
The final solution included weights for each plate based
on the measuring accuracy of the reference stars of that particular plate.
Our catalog includes mean-epoch positions (1975.0) and proper motions
for 2272 objects. The photometry is that derived in Section 3, where
for stars in common with S89 we have replaced our
photographic photometry with that in S89. The internal error of the
proper motions is a function of magnitude, and this is represented in
Figure 1, top panel. Well-measured stars ($V \le 18$) 
have a mean internal positional error of 
0.3 $\mu$m (3.3 mas) at the mean epoch 1975.0, and a
mean internal proper-motion error of 0.3 mas yr$^{-1}$.
The lower panels of Fig. 1 show proper motions in the  
$x$ and $y$ coordinates respectively, versus magnitude. The filled triangles
represent the photometrically-selected blue stragglers from 
R98, while the filled circles represent 
four bright red giants that are cluster members based on radial velocities
(Da Costa \& Armandroff 1991). 
Since we had few cluster stars to define it confidently,
a small amount of magnitude equation is left at the very bright end
in $\mu_{x}$.
The red clump stars of the cluster ($V = 17$, $B-V = 0.75$, see
S89, can be detected in the lower panels of Fig. 1
 as a slight over density at $V = 17$. In these 
plots, the location of the red clump stars compared to that of 
the blue stragglers shows that we have obtained 
proper-motions free of systematics over a significant magnitude and
color range. 

Proper-motion membership probabilities were calculated from 
the proper-motion distribution in a manner similar to that in
Dinescu {\it et al.} (1996). We defer the discussion of the 
membership probabilities in relation to the CMD and
cluster structural parameters to a future paper.


\section{The Absolute Proper Motion of Pal 12}

\subsection{The Mean Motion of the Galaxies}

We have a list of 121 galaxies for potential use 
in the correction to absolute proper motion for Pal 12.
In order to derive the absolute proper motion,
one can, in principal, apply as an offset the mean motion of galaxies 
to the mean motion of clusters stars, where both these means 
are readily estimated from the relative proper motions 
as derived in Section 4. However, the relative proper 
motions derived in Section 4 are based on a global solution, 
in which we hope to best model the geometric systematics. 
While the plate model may be represented accurately in a small region 
around the plate center, where
there are plenty of stars and geometric systematics are small,
this is not the case as one moves away from the plate center.
Specifically, proper motions are affected by unaccounted for
geometric systematics towards the outer regions of the field, and it is
in this region where most of the galaxies lie. If we were to limit the
survey 
area to a radius of $30$ mm ($5\farcm5$) from the plate center,
we would have only 4 galaxies to establish the absolute reference frame.

In order to include the entire list of galaxies in our solution
we have performed a local solution as developed by 
Dinescu {\it et al.} (1997), which has the advantage of eliminating 
geometric systematics at the expense of introducing more random 
noise that comes from the intrinsic proper-motion dispersion of the
field stars. Thus, under the assumption that geometric systematics 
affect all objects in a small region of the plate in the same way, we
rederive the proper motion of each galaxy locally. This ``local''
proper motion  is the difference between 
the proper motion of the galaxy obtained from the global solution 
(described in Section 4), and the mean proper motion of a local 
reference system of stars surrounding the galaxy. 
The proper motions of the stars that make the local reference system
also come from the global solution. 
In practice we use the median proper motion as it is less sensitive to
outliers, especially important when the 
the local system consists of 
a few (up to ten) stars. With the described methodology
 the geometric systematics 
are cancelled out locally. However, the actual stellar population that 
defines the local reference system is different from the stellar
population that defines the reference system in the global solution. 
Thus, in the global solution the reference system consists of a 
combination of cluster and field stars in a 
relatively large magnitude range, while the local solution, which
uses the stars in the vicinity of each galaxy, may have no cluster
stars. In other words, there is no guarantee that each local reference
system has the same ratio of cluster stars to field stars as
the global solution does, and, consequently, that they are equivalent.
Therefore, in order to tie the mean motion of the
cluster to this locally-derived mean motion of galaxies, we have to
know exactly to which population of stars the locally-defined 
reference systems belong to, and to rederive the mean motion of the 
cluster with respect to this population, rather than that of the combined
reference system used in the global solution.

We have 
selected field stars to comprise the local-solution reference samples
by including stars with proper-motion cluster membership probability 
$\le 30\%$ (see Section 4), in the magnitude range $V = 17$ to 21.5.
Kinematic models of the Galaxy (M\'{e}ndez 1995)
 show that, in this magnitude range,
the mean motion of the field (secular proper motion) changes slowly
with magnitude (less than 0.5 mas yr$^{-1}$), and the intrinsic
dispersion of the field also becomes smaller ($\sim 5 - 7$ mas yr$^{-1}$)
than at brighter magnitudes.   Both these trends work to minimize 
the scatter in the mean motion of the local system.
After several experiments with the number of stars that define the
local solution, we have chosen seven stars per galaxy, 
and these were each also required to
have at least 8 plate measures and a proper-motion error of less than 
4 mas yr$^{-1}$ in each coordinate. The average radius of these local
systems is $\sim 5.6$ mm. This radius is set such that
the shift produced by the largest, highest-order term
in our global plate solution over this area is less than the 
typical positional measuring error of our stars
(e.g., over a size of $2 \times 5.6$ mm, 
the largest 4th-order term produces a 
displacement of 2.2 mas; see also Section 4).
In Figure 2 we show the run of proper motions of galaxies 
versus positions, magnitudes and colors. The left-side panels
represent the proper motion along the $x$ coordinate and the right-side
panels the proper motion along the $y$ coordinate. The top row panels
show the proper motions obtained from the global solution
(cross symbols), versus coordinate (either $x$ or $y$ coordinate, whichever plot showed more prominent systematic trends with position). 
The next row of panels
shows the proper motions obtained from the local solution
(filled circle symbols); as expected 
the positional trends are diminished. 
The last four panels show the local-solution proper motions 
as a function of magnitude and 
color, and no significant trends are seen in these plots.

The mean motion of galaxies with respect to this reference system is:
$\mu_{G,x}^{R'} = -1.27\pm0.30 $ mas yr$^{-1}$, and $\mu_{G,y}^{R'} = 5.49 \pm 0.29
$ mas yr$^{-1}$. 
Since we believe that most of the systematics in the proper motions
are eliminated to the extent this plate material and our techniques 
allow it, we have  
calculated this value as a weighted mean, where the 
weights are given by the estimated proper-motion uncertainty of each galaxy. 
Galaxies that had proper motions that differed by more than $2.5 \times$ the
standard deviation from the mean were eliminated in an iterative selection.
The uncertainty
in the average is calculated based on the scatter about the
average and the weights, and it will remain the dominant source of error
in our final absolute proper motion of Pal 12.

\subsection{The Mean Motion of the Cluster}

The mean motion of the cluster was determined from the fit of a
model to the 
proper-motion distribution in each coordinate. The proper-motion distribution
was constructed from the set of discrete proper motions, by smoothing the data
with a Gaussian of width equal to the proper-motion uncertainty of each star
(see formula 1
in Dinescu {\it et al.} 1996). The model fitted consists of the sum of two
Gaussians, which represent the cluster and the field distribution. 
In this case, when we are mainly concerned with accurately determining 
the mean motion of the cluster, we have restricted our 
observed proper-motion distribution to a magnitude range
$V = 16$ to 21, a radius from the plate center of $30$ mm ($5\farcm46$), 
and a
proper motion interval $| \mu_{x,y}| \le 15$ mas yr$^{-1}$. 
The magnitude and radius restrictions assure that the sample has the
best-measured, least-prone-to-systematics proper motions,  while the 
proper-motion restriction has the role of obtaining a good fit in the 
vicinity of the cluster peak. We have thus obtained a 
mean motion of the cluster with respect to the reference stars:
$\mu_{C,x}^{R} = -1.34\pm0.01 $ mas yr$^{-1}$, and $\mu_{C,y}^{R} = 
1.08 \pm 0.01
$ mas yr$^{-1}$, where the uncertainties represent the formal uncertainties
 from the fit. A more realistic value of the uncertainty is of the 
order of 0.05 mas 
yr$^{-1}$, which is determined from the width of the Gaussian distribution 
of cluster stars, and the number of cluster stars, as
derived from the fit. 

\subsection{Final Absolute Proper Motion}

One more step is necessary in order to bring to absolute  
the mean cluster motion derived in the previous subsection. We must 
determine the difference between the mean motion of the 
reference system used in the
global solution of Section 4, and that of the local reference system
derived in Section 5.1. To do so, we select the stars
with the properties of those stars used in the local solution
($V = 17 - 21.5$, membership probabilities less than 30\%, 
number of plate measurements $\ge 8$, 
proper-motion errors $\le 4$ mas yr$^{-1}$),
and located within a radius of 30 mm from the plate center
--- which is the area where we have determined the mean motion of the
cluster. Then
we calculate the median of this sample: $\mu_{x}^{R'-R} = 1.13$ mas yr$^{-1}$,
and $\mu_{y}^{R'-R} = -0.20$ mas yr$^{-1}$.
The final absolute proper motion of the cluster is given by
$\mu_C = \mu_C^{R} - \mu^{R'-R} - \mu_G^{R'}$ in each coordinate, with
the result:
$\mu_x = - 1.20 \pm0.30$ mas yr$^{-1}$, and $\mu_y = -4.21 \pm0.29$ mas 
yr$^{-1}$. The uncertainties represent the square root of the quadrature sum
of the error in the mean motion of galaxies (Section 5.1), 
and that of the mean motion of the cluster (Section 5.2).
Since the right ascension and declination are aligned with the $x$
and $y$ coordinate respectively, this proper motion can be regarded as that
in the direction of right ascension and declination respectively.

\section{The Orbit of Pal 12: Another Sgr Cluster ?}

\subsection{Space Velocities}

We have adopted a standard solar motion of 
$(U_{\odot}, V_{\odot}, W_{\odot}) = (-11.0, 14.0, 7.5)$ km s$^{-1}$
(Ratnatunga, Bahcall \& Casertano 1989) with respect to the Local
Standard of Rest (LSR). Here the $U$ component is positive outward from the
Galactic center (GC). The adopted rotation velocity of the LSR is 
$\Theta_{0} = 220.0 $ km s$^{-1}$, and the solar circle radius is
8.0 kpc. The adopted distance to Pal 12 is $19.5 \pm 0.9$ kpc
(R98), the heliocentric
radial velocity is $27.8 \pm 1.5$ km s$^{-1}$, and the Galactic
coordinates are: l = $30\fdg512$, b = $-47\fdg681$ (Harris 1997).
The derived LSR velocity is $(U, V, W) = (-225\pm24, -329\pm30, -21\pm19)$
 km s$^{-1}$, 
and the corresponding velocity in a cylindrical coordinate
system centered on the Galactic Center is $(\Pi,\Theta, W) = (2\pm29, 250\pm25, -21\pm19) $ km s$^{-1}$.

\subsection{Orbital Elements}

In order to obtain the orbital elements, we have integrated the 
orbit of Pal 12 in a three-component, analytical model of the
Galactic gravitational potential.
The bulge is represented by a Plummer potential, the disk by a 
Miyamoto \& Nagai (1975) potential, and the dark halo has a logarithmic
form. For the exact form of the potential see 
Paczy\'{n}ski (1990).

The orbital elements were calculated as in Dinescu {\it et al.} (1999).
They are averages over a 10-Gyr time interval. The uncertainties 
in the orbital elements
were derived from the width of the distributions of orbital elements,
over repeated integrations, which had different initial positions and
velocities. These positions and velocities were derived in a Monte Carlo 
fashion from the uncertainties in the observed quantities:
proper motions, distance and radial velocity. For details we 
refer again the reader to Dinescu {\it et al.} (1999). We thus obtain 
an orbit of pericentric radius $R_{p} = 16.0\pm0.6$ kpc, 
apocentric radius $R_{a} = 29.4\pm6.0$ kpc, maximum distance above the 
Galactic plane $z_{max} = 20.1\pm2.5$ kpc, eccentricity $e = 0.29\pm0.08$, and
inclination with respect to the Galactic plane $\Psi = 58\fdg3 \pm 2\fdg2$.
The azimuthal period is $P_{\varphi} = (0.73 \pm 0.11) \times 10^{9}$ yr.
With the present location of Pal 12 at a distance of $\sim 16.2$ kpc
from the GC, this implies that the cluster is at its pericenter.

\subsection{Comparison with the Orbits of Sgr and LMC}

Since the work of LR92, there has been the suggestion of an association of
Pal 12 with the LMC.
The main argument prompting LR92 to investigate this hypothesis was the
young age of Pal 12 compared to that of traditional halo globular
clusters of similar metallicity (see for instance S89).
LR92 explored possible orbits for two young globular clusters --- Pal 12 
and Rup 106 --- under the assumption that they were captured from the LMC. 
The main observational constraint for the calculated family of orbits
was the radial velocity of the cluster,  and LR92 concluded that
Rup 106 is more likely to have been torn from the LMC than Pal 12. 
Nevertheless, they predicted a transverse motion 
for Pal 12 under the assumption that the cluster is currently
at its pericenter (inferred from its small radial velocity)
and that its apocenter
should be at least of the size of the pericenter of LMC's orbit 
(i.e. $\sim 50$ kpc).  The predicted Galactocentric transverse motion 
is  $\approx 3.5$ mas yr$^{-1}$ towards the LMC plane (LR92).

Armed with the proper motions of Pal 12, we investigate here 
the motion of Pal 12 in relation to the LMC.
Subtracting the Solar peculiar motion and the Galactic rotation from the 
just-derived proper motion of Pal 12 (Section 5.3) we obtain the 
transverse motion in Galactic
coordinates: $\mu_l~cos~b = -2.26\pm0.30$ mas yr$^{-1}$ and 
$\mu_b = 0.95\pm0.30$ mas yr$^{-1}$ (see also Table 3).
This gives a total transverse motion of $2.45 \pm 0.3$ mas yr$^{-1}$, at
an angle of $23^{+10}_{-9}$ degrees with respect to the line of Galactic 
latitude that goes through Pal 12, in the direction of antirotation. 
The angle between the line of latitude that goes through Pal 12, and
the great circle that goes through Pal 12 and the LMC 
($l = 280\fdg5, b = -32\fdg9$) is $\chi_{LMC} \sim 16\arcdeg$. 
Therefore the proper motion of Pal 12 appears to be oriented
toward the LMC within the uncertainties.

Using our velocity components, $\Pi$, $\Theta$ and W (Section 6.1), we obtain
a Galactocentric proper motion of $3.25$ mas yr$^{-1}$, and
a Galactocentric radial velocity of 21 km s$^{-1}$. The
Galactocentric  transverse motion we obtain does indeed agree
reasonably with the 3.5 mas yr$^{-1}$ value predicted by LR92. 
However, we note that in their
calculation LR92 made the approximation that the value of the line-of-sight
velocity equals that of the Galactocentric radial velocity.
This approximation is not valid for the case of Pal 12, which
has a heliocentric radial velocity of 107 km s$^{-1}$ 
(in a Galactic rest frame; see 
also LR92), a value that is much larger than the actual Galactocentric
radial velocity calculated from the full velocity vector.

A second candidate for a Pal 12  parent galaxy is 
Sgr, discovered in 1994 by Ibata, Gilmore \& 
Irwin. A first indication of a connection is
the location of Pal 12 with respect to Sgr:
they are relatively close in Galactic longitude 
($l_{Pal~12} = 30\fdg5,~l_{Sgr} = 5\fdg6$), 
and have similar Galactocentric radii 
($R_{GC}^{Pal~12} = 16.2$ kpc,~$R_{GC}^{Sgr} = 17.4$ kpc). 
The angle between the line of Galactic latitude that goes through 
Pal 12, and the great circle that goes through Pal 12 and Sgr is
$\chi_{Sgr} = 53\arcdeg - 58\arcdeg$, where the range is given by 
the extended size of Sgr. Here we have taken the range 
as given by clusters M 54
and Arp 2, which are associated with Sgr (Da Costa \& Armandroff 1995). 
Thus, the proper motion of Pal 12 is not aligned with the great circle 
that goes through Sgr and Pal 12. However, it can be considered 
``toward'' Sgr in the sense that the angle difference 
between the great circle that goes through Sgr and Pal 12,
and the direction of the proper motion is less than $\sim 30\arcdeg$.

At this point, solely from the proper motion of Pal 12 and the location 
of the two candidate host galaxies, one can not discriminate
against/for the host galaxy. This is because 
the kinematics can be relatively easily interpreted only 
in the case the disruption event took place recently.
Otherwise, effects such as the precession of the orbits,
and wrapping around of the tidal streams for instance,
 can easily complicate the kinematical interpretation. 
That the presumed tidal stripping did not occur recently, at least from the
LMC, one can
intuitively see from the fact that the cluster is not close to 
the orbital plane of the LMC.
We remind the reader that the LMC plane is defined by the Magellanic clouds,
Ursa minor, Draco, the Magellanic stream and the motion of the LMC
(Lynden-Bell \& Lynden-Bell 1995).

Finally, even though our proper motion measurement happens to match
well the prediction 
made by LR92, it is not necessarily for the right reason, and, in any case,
such a match, were it real, would not in itself provide conclusive
evidence of association.

Therefore we proceed to look at the orbital elements for a better
understanding of the present kinematical data. 
The values of the proper motions in equatorial coordinates, 
Galactic coordinates, and Galactic coordinates with the Solar motion and
Galactic rotation subtracted, are summarized in Table 3 for Pal 12,
Sgr and LMC. The symbols Sgr1 and Sgr2 in Table 3, and throughout the 
text will refer to the two proper-motion measurements of Sgr.
The two determinations of Sgr's proper motion are:
the one derived by Irwin {\it et al.} (1996) from Schmidt 
plates (Sgr1), and the one derived by Ibata {\it et al.} (1998) from HST WFPC2
frames (Sgr2). 

We have also determined the orbits, orbital parameters and their 
uncertainties for Sgr and the LMC. 
These are summarized in Table 4
(total orbital energy $E_{orb}$, orbital angular momentum $L_{z}$,
total angular momentum $L$, azimuthal period $P_{\varphi}$,
radial period $P_r$, apocenter radius $R_a$, pricenter radius
$R_p,$ maximum distance above the Galactic plane $z_{max}$, 
eccentricity e, and inclination of the orbit with respect to the
Galactic plane $\Psi$), along with those for Pal 12.
We have adopted a distance to Sgr of $25\pm2.5$ kpc, and 
a heliocentric radial velocity of $137 \pm 5$ km s$^{-1}$ (Ibata {\it et al.}
1997). 
The proper motion for the LMC is an average of three studies:
Jones, Klemola, \& Lin (1994), Kroupa, R$\ddot{o}$der, \& Bastian (1994),
and Kroupa, \& Bastian (1997). The adopted distance from Sun to  the
LMC is $49\pm5$ kpc, and the heliocentric radial 
velocity is $270\pm4$ km s$^{-1}$ (see, e.g., Kroupa \& Bastian 1997,
Meatheringham {\it et al.} 1988).

From Table 4, it is clear that the LMC's orbit is significantly more
energetic than that of Pal 12. The apocenter  of 
Pal 12 is smaller than the pericenter of LMC at a $1.2~\sigma$ level.
In principal, tidal debris
has an orbital energy that is close to that of the satellite that is
disrupted (Johnston 1998 and references therein). 
Thus, following Johnston (1998), the amount
of change in the total orbital energy (or the amount of orbital energy
lost/gained due to the tidal interaction) is given by the gradient in the
gravitational potential of our Galaxy over the size of the satellite, and
is of the order of $(\frac{M_{sat}}{M})^{1/3} \times E_{orb}$, where 
$M_{sat}$ is the
mass of the satellite, and $M$ is the mass enclosed within the satellite
orbit. For a mass of the LMC in the range of $10^{9}$ to $10^{10}$ M$_{\odot}$,
and a mass of the Galaxy enclosed within LMC's orbit of 
$\sim 5.0 \times 10^{11}$ to $10^{12}$
 M$_{\odot}$, we obtain that the tidal debris from LMC should have
orbital energies that do not vary by more than 10 to $27\%$ from LMC's
orbital energy. If Pal 12 were torn from the LMC, then the cluster lost
$\sim 36 \%$ of the total orbital energy of the LMC (Table 4), a value that 
is  larger than the range predicted by simple estimations.

Inspecting the orbital parameters of Pal 12 and Sgr
 one can see that the orbits are relatively similar. 
Pal 12 has a $19.8\%$ less energetic orbit than Sgr, and consequently its 
apocentric distance and orbital eccentricity are smaller than those
 of Sgr. Their pericentric radii and inclinations with respect to the
Galactic plane agree very well. The orbital angular momentum
$L_{z}$ seems to be somewhat discrepant, such that
Pal 12 appears to have more rotation than Sgr.
However, for highly inclined orbits such as those derived for the
two objects,
the orbital angular momentum is not the most appropriate 
quantity to characterize the orbit, in spite of 
its characteristic as an integral of motion.
For example, the second determination of Sgr's orbit shows 
more rotation because the proper motion is slightly different than that of the
first determination (see Table 3); at distances from the Sun of  
$\sim 25$ kpc, such small changes in the proper motion for highly-inclined
orbits can alter  the orbital angular momentum significantly.
A more appropriate quantity is the total angular momentum, which
can be regarded as a third integral of motion, especially for 
objects in the outer halo, where the gravitational potential
becomes more spherical (Binney \& Tremaine 1987).
 We have specified this quantity which represents the average over the
10-Gyr integration time, in the third column  of Table 4.
Again, Pal 12's total angular momentum is quite similar to that 
of Sgr, but significantly smaller than that of LMC. 

Since satellite disruption models show that tidal debris
have orbits that energetically resemble closely that of the original satellite,
 it seems that Pal 12's origin 
as a Sgr cluster represents a much more feasible scenario than
that of Pal 12 originating in the LMC, and subsequently 
undertaking a significant energy and angular momentum loss.
Therefore we will further explore this scenario in more detail.

\section{Exploring the Origin of Pal 12 as a Sgr Cluster}

\subsection{Phase-Space Coincidence}

In order to assess whether it is likely that Pal 12 and Sgr
have a common origin, besides the simple comparison of their orbital
elements, we 
have integrated the orbits back in time, using the same, constant
time step ($10^{5}$ yr), and compared the positions and 
velocities of the two orbits at each step.   
We have looked at two quantities that can represent a 
boundness criterion.
The first one is a normalized distance in phase space: 
\begin{equation}
d^{2}  = \frac{\Delta r^{2}}{a^{2}}+ \frac{\Delta V^2}{b^{2}}
\end{equation}
Here $\Delta r^2 = \sum_{i=1}^{3}(x_{1}^{i}-x_{2}^{i})^2$, 
$\Delta V^2 = \sum_{i=1}^{3}(v_{1}^{i}-v_{2}^{i})^2$,
and $x$ represents the spatial component, $v$ is the velocity,
and the indices 1 and 2 stand for the orbit of Pal 12 and Sgr
respectively. The constant $a$  represents the spatial scale over
which we wish to find coincidence, and it is given by the tidal 
radius of the satellite, while $b$ is the velocity size over which 
we want to find coincidence, and it is given by the escape velocity
from the  satellite. These two constants have the role of 
normalizing our ``distances'' in real physical spaces, such that
we can construct a non-dimensional phase-space distance that constrains at the
same time the spatial and velocity ``distances'', and
has the property of  physically representing the limit of
boundness to the parent satellite. We have chosen 
$a = 3$ kpc, and $b = 38$ km s$^{-1}$. These numbers are
representative for a satellite of mass $10^{8}$ M$_{\odot}$
and effective radius  600 pc, which are
characteristic of a Fornax-like satellite 
(see e.g., Irwin \& Hatzidimitriou 1995).
We have chosen this value for the mass from the wide range in published 
Sgr mass estimates: from $10^{7}$ to $10^{9}$
M$_{\odot}$ (see discussion in Ibata {\it et al.} 1997). 
The escape velocity is estimated at the effective radius,
using formula (2.26) in Binney \& Tremaine (1987).
The choice of $a = 3$ kpc may not be consistent with that of 
$b = 38$ km s$^{-1}$, but we argue that these values are representative 
for the disruption event we want to detect.
Thus, for our example satellite, the cluster needed an escape velocity of
38 km s$^{-1}$, but it need not necessarily be at a radius of 600 pc from
the satellite. Shortly after the cluster escaped the potential field
of the satellite, it still comoves with the satellite in the vicinity of the
satellite. We have chosen the size of this vicinity to be 3 kpc. This is 
not unreasonable; for instance, the distances between M54 (usually chosen as
the center of Sgr) and the other three clusters associated with Sgr,
Ter 7, Arp 2 and Ter 8 are respectively: 4.2, 3.6 and 4.8 kpc (Harris 1996).
We also note that, a presumably more appropriate, larger mass
of the satellite, of the order of $10^{9}$ M$_{\odot}$ would require
a larger escape velocity, which would make our criteria less conservative
than our current choice.


For our case of study, if the distance in phase space is equal to or
less than 1.41, it implies that the corresponding orbit-combination is the
one in which Pal 12 and Sgr had a common origin.

The second boundness criterion is less restrictive in terms of the
properties of the parent satellite, but consistent with respect to
the escape velocity and the radius where the escape event takes place. The 
quantity $\Delta r~\Delta V^{2}$ is calculated and is directly
compared to $2 G M_{sat}$, where $G$ is the constant of gravitation and 
$M_{sat}$ is the mass of the satellite within the radius $\Delta r$.
This formulation handles better the 
dependence of the escape velocity with the radius,
as it reflects the energy balance for the escaping cluster, however 
it has the property of over weighting the $\Delta V^{2}$ quantity in the
criterion. Therefore this criterion should be regarded in conjunction 
with our first criterion.

For a given $M_{sat}$, if $\Delta r~\Delta V^{2} \le  2 GM_{sat}$,
the cluster can be considered bound to the satellite at that 
particular time. 

Figure 3 shows $\Delta r$, $\Delta V$, $d$, and $\Delta r~\Delta V^{2}$
in each panel respectively as a function of time. We have restricted our
plot to $-4$ Gyr (the minus indicates a backward time integration),
 as we have found that for larger integration times, the orbits 
tend to diverge. 
We have considered four cases corresponding to the two orbits 
of Sgr (Table 3), and two
Galactic potential models. One model is that used in 
Section 6 (Paczy\'{n}ski 1990, P90 model), the other is that defined in
Johnston {\it et al.} (1995) (JSH95 model). 
The P90 model has a smaller, more centrally concentrated bulge than
JSH95, and also a less massive, less flattened disk than the JSH95 model.
The dark halo has a logarithmic form and it is spherically symmetric
for both potentials. Our preferred model is P90 (see Dinescu {it et al.} 1999b)
as it is  more realistic than the JSH95 model
in terms of disk and bulge shocking for the globular clusters. It also 
provides slightly longer periods, as the orbits are more energetic,
and, in the inner regions, has a different precession rate than the
JSH95 model.

The thick lines represent the Sgr2 orbit, while the thin lines
represent the Sgr1 orbit (Table 3 and 4). The continuous lines correspond to
 the P90 model, while the dashed lines to the JSH95 model.
From Figure 3 the minimum $d$ occurs at $t \sim -1.7$ Gyr for both 
Sgr1 and Sgr2 orbits in model P90, while for model JSH95, the 
minimum $d$ occurs at $t \sim -1.4$ Gyr, also for both orbits of Sgr.
For interpretation of minima in 
the quantity $\Delta r~\Delta V^{2}$ one should 
also regard simultaneously $\Delta r$, as it is possible to find
minima of $\Delta r~\Delta V^{2}$ that correspond to a coincidental 
situation in which Pal 12 and Sgr have relatively close
velocities, but are located at too large of a distance from one another to be 
physically bound. Such a  specious minima are those
at $t \sim -3.4$ Gyr for model 
P90 and $t \sim -2.7$ Gyr for model JSH95, where each corresponds to a  
$\Delta r$ of the order of 50 kpc. Interestingly, at the minimum
of $d$, we also find the second minimum for $\Delta r~\Delta V^{2}$,
and this is valid for both models and Sgr1 and Sgr2 orbits. 
Technically, if indeed this point describes the moment when Pal 12
was last bound to Sgr, then the rest of the points at earlier times
have no physical significance, 
as our backward integration continues to
treat the process as a simple Galaxy-satellite problem, 
while the reality is an interaction between the Galaxy and at least 
a two-body system.
Among the four model-proper-motion combinations 
in Figure 3, the smallest minimum  
in both $d$ and $\Delta r~\Delta V^{2}$ corresponds 
to the Sgr2 orbit in model P90. The values are:
$d_{min} = 4.7$, $\Delta r_{d_{min}} = 11$ kpc, and 
$\Delta V_{d_{min}} = 113$ km s$^{-1}$ at $t = -1.69$ Gyr; 
and log$_{10}(\Delta r~\Delta V^{2})_{min} = 5.0$,
$\Delta r_{(\Delta r~\Delta V^2)_{min}} = 15 $ kpc, and
$\Delta V_{(\Delta r~\Delta V^2)_{min}} = 81 $ km s$^{-1}$, at
$t = -1.58$ Gyr. 
These numbers are to be compared with 1.41 for $d_{min}$ and, 
 log$_{10}(\Delta r~\Delta V^{2}) =$
1.94, 2.94, and 3.94 for $M_{sat} = 10^{7}, 10^{8}$ and $10^{9}$
M$_{\odot}$ respectively. Therefore, at a first inspection,
our criteria for boundness are not satisfied. 
However we note the relatively large width of this minimum, in both
quantities. This indicates that, for a considerable amount of 
time, the two orbits are in phase, or the objects are moving
together, occupying the same region in phase-space.
Also, since the orbit of Pal 12 is 19.8\% less energetic than that of 
Sgr (Table 4, Section 6.3), we can not expect to satisfy strictly
the energy balance $\frac{\Delta V^2}{2} = \frac{G~M_{sat}}{\Delta r}$,
but one should take into account the difference in kinetic energy, 
due to the fact that the 
two orbits are energetically different at a 19.8\% level. While rigourosly
incorporating this energy difference in our boundness criteria requires
simulations that we will address later in this Section, here we estimate, for
illustration purposes, how the limit in $\Delta r~\Delta V^{2}$ can change.
If we assume that, at the desired minimum, Pal 12 and Sgr are spatially
located close to each other, such that their potential energy is practically
the same, then the difference in kinetic energy should be equal to the 
difference in total orbital energy. From Table 4, this corresponds 
to $\Delta E_{kin} = 1.6 \times 10^{4}$ km$^2$ s$^{-2}$. For a reference 
$\Delta r = 3$ kpc the above-mentioned limits in
log$_{10}(\Delta r~\Delta V^{2})$ become 4.98, 4.99, and
5.02 for the three values of Sgr's mass. Therefore the values that we have
obtained for Pal 12 and Sgr2 should be regarded in the context 
of the orbital-energy difference between the two objects.

In addition to these arguments, the uncertainties
in the measured orbits will provide a range for the 
quantities that define our boundness criteria.
To asses the size of the uncertainty
in our derived $d_{min}$ due to the uncertainties in the orbits
we have calculated $d_{min}$ for a set of orbit combinations.
Using a Gaussian representation
of the uncertainties in the proper motions, radial velocities and distances, 
we have derived 100 orbits for Pal 12 and Sgr2, and
have therefore $10^4$ orbit combinations. The representative $\sigma$ 
of each observable is taken from Table 3, and Section 5.3 and 6. 
This simulation, besides being representative for
observational uncertainties, can also be regarded as an exploration
of our boundness criteria as a function of the orbital difference
between Pal 12 and Sgr. Thus, in
Figure 4, left panel we show  the relative difference between the orbital
energy of Pal 12 and Sgr as a function of $d_{min}$.
The negative sign in the ordinate
means that Pal 12's orbit is less energetic than that of Sgr, and,
from the standpoint of tidal streams, it corresponds to leading streams.
There are only 15 orbit combinations of the $10^{4}$ explored
that have $d_{min} \le 1.41$ and therefore strictly satisfy our criterion. The
orbital-energy difference for these is no larger than 4 \%, and they cluster
at two epochs: -1.7 Gyr and -0.9 Gyr.  At progressively larger 
orbital-energy differences, $d_{min}$ increases as it is intuitively
expected, since the limit of $d_{min} \le 1.41$ is designed for
orbits of very similar total orbital energy. This value 
can be no longer considered the boundness limit in a strict sense, and 
the left panel of Figure 4 has the role of showing the range of this quantity
for a particular orbital-energy difference. 
Since the conservatively-defined boundness criterion ($d_{min} \le 1.41$) was
achieved in this total family of orbits for orbits that differ by no more than $4\%$ in their orbital energy, one can assume that a bound case can be achieved for orbits that have progressively higher orbital-energy difference.
Thus the lowest $d_{min}$ for a given range of the orbital-energy difference
can be regarded as the ``modified-$d_{min}$'' criterion.

If from the $10^4$ orbit combinations we select only those  
that have an orbital-energy difference between -22\% and -18\% 
(range which closely brackets the value for the Pal 12 - Sag2 case), 
and we plot
their $d_{min}$ as a function of time (right panel of Figure 4) we can
see that this restricted population
has two preferential moments of low $d_{min}$; one at $t \sim -1.5$,
and another one at $t \sim -0.1$. The former minimum also corresponds
to the 
lowest value; $d_{min} = 2.9$, and $\Delta r_{d_{min}} = 6$ kpc,
$\Delta V_{d_{min}} = 76$ km s$^{-1}$, and 
log$_{10}(\Delta r~\Delta V^{2})_{d_{min}} = 4.5$, and it is adopted 
as the boundness limit for this restricted range 
in the orbital-energy difference.
The uncertainties in these quantities are taken as the standard
deviations of this population of orbit-combinations,
and they should be regarded as lower limits, since we restricted the
range in the orbital-energy difference. They are:
$\sigma_{d_{min}} = 1.5$, 
$\sigma_{\Delta r_{d_{min}}} = 4$ kpc,
$\sigma_{\Delta V_{d_{min}}} = 55$ km s$^{-1}$, and
$\sigma_{log_{10}(\Delta r~\Delta V^{2})_{d_{min}}} = 0.3$.
If we consider these uncertainties, and if we adopt the lowest 
$d_{min}$ as the boundness limit, which, incidentally, occurs 
at a moment very close to the moment of our measured case, then 
we can state that the measured case is within $1\sigma$ of 
a case of a bound Pal12-Sgr2 system at $t \sim -1.7$ Gyr.

The uncertainties in the orbits are relatively large,
especially if we use the conservative proper-motion error for Sgr of
0.8 mas yr$^{-1}$ (Table 3). We realize that, with these uncertainties,
a large range in the properties of the orbit-combinations is obtained, 
and therefore, for those orbit-combinations in which the capture events occur,
the timing may be different, and thus not well constrained.
We have shown however that the epoch  $t \sim -1.7$ Gyr
appears in cases within the boundness limit for at least two 
values of the orbital-energy differences. 

The same exercise can be repeated for the 
Sgr1 proper motion with its quoted error which is 10 times 
smaller that the value we have used here. Most likely it would lower the
probability for the bound case  because of the larger $d_{min}$ value
than that for Sgr2 (see Figure 3). The energy spectrum would be better 
constrained, and therefore most of the orbit combinations
would lie closer to the $\Delta E_{orb} = -19.8 \%$. The uncertainties 
we derived for the Sgr2 can still be regarded as representative as
they were determined in a narrow $\Delta E_{orb}$ range. What is
uncertain is the timing of the capture event: such simulations may 
not be able to distinguish preference between
the $t \sim -1.7$ Gyr and 
the $t \sim -0.1$ Gyr. This is seen in the third-from-the-top panel of
Figure 3: for the Sgr1 (thin, continuous line) the two values of $d$ at
the respective epochs are closer in their value than those for Sgr2.

\subsection{Tidal Streams from Sgr}

With various studies of the Sgr population at large distances
from the main body of the dwarf galaxy ($l \sim 6\arcdeg,~
b \sim -14\arcdeg$, Ibata {\it et al.} 1997), we 
focus our attention on the work of Mateo, Olszewski \& Morrison (1998)
and Majewski {\it et al.} (1999) (hereafter M99). 
Mateo, Olszewski \& Morrison (1998) report a Sgr population
in excess of the background as far from the core as 
$l \sim 9$ and $b \sim -46$,
while M99 find evidence for Sgr at
$l \sim 11\arcdeg$ and $b \sim -40\arcdeg$. It seems that the dwarf galaxy,
or at least part of it, extends toward greater longitudes
as it extends toward lower latitudes.

Given the widely-accepted proper motion of Sgr (Table 3), which has 
little or practically no motion along Galactic longitude, most 
Sgr starcount surveys are focused along a narrow longitudinal band.
It is worth mentioning that at least
one of the proper-motion determinations of Sgr
(Irwin {\it et al.} 1996), is poorly constrained in longitude, since
the absolute reference system did not consist of background galaxies,
but rather stars with supposedly known kinematics
(red giants in the disk). 

With these considerations in mind, we investigate the 
possibility that Pal 12 is part of a tidal stream from Sgr.
This investigation follows closely that of J99.
The J99 work presented a comparison between 
observations of the Sgr population (location, surface density
and radial velocity) and the predictions of a semi-analytic model
of the phase-space structure of the tidal debris. J99 showed that 
the low-latitude detections ($b \sim -40\arcdeg$) can be plausibly explained
by a leading stream that was lost from Sgr at a 
pericenter passage $n_{p} = -2$, where $n_{p} =0$ is the present
pericenter passage. There were three main observational constraints:
the distance and the radial velocity of the candidate Sgr red clump stars
identified in the study of M99 that matched the 
predictions for the leading stream, and the break found in the
surface density at $b \sim -30\arcdeg$ (see Fig. 4 in J99 and 
Fig. 3 in Mateo, Olzewski \& Morrison 1998) which makes a transition
between the main body of Sgr and the leading stream.

Here we have reinvestigated the problem, considering 
Pal 12's motion.  We have considered only our preferred Sgr 
proper motion (see Sections 6.3, 7.1), which is Sgr2 in Table 3.
Figure 5 shows the distribution and the kinematics of the tidal 
streams compared to those of Pal 12. 
In all panels Sgr is represented by the filled 
triangle, Pal 12 by the filled circle, and the candidate
Sgr red clump stars from
M99 by the filled square. All leading streams
are represented with smaller filled circles, while the trailing streams
are represented with open circles. As in J99, the darkest shade represents
the most recent stream ($n_{p} = 0$), and progressively lighter 
shades represent accordingly previous pericentric passages. 
The top two panels show the present spatial distribution of the streams,
and of the points in discussion; Pal 12 and M99. 
On these plots we have also superimposed the orbits.
The orbit of Sgr is represented 
with a continuous line, and that of Pal 12 with a dashed line.
In these coordinates the Galactic center is at (X,Y,Z) = (0,0,0) kpc
and the Sun at (X,Y,Z) = (8.0,0,0) kpc.
The following panels show the velocities in a Galactic reference frame 
along the line-of-sight,  Galactic longitude and Galactic
latitude respectively as a function of Galactic latitude.
The last panel shows the 
relative orbital-energy change with respect to that of Sgr
as a function of latitude. These last four panels represent a subsample of 
data points in longitude, ranging from $340\arcdeg$ to $70\arcdeg$.
This was done in order to avoid confusion in the kinematics of the
particular streams in discussion. As in J99, the stream suspected 
to be represented by M99 and, here, by Pal 12 as well, is 
the leading stream for $n_{p} = -2$; in our plots it is the lightest 
shade of filled circles. 

From the first two panels which show the
spatial distribution, we can see that the location of both
Pal 12 and M99 follows closely that of the $(n_{p}=-2)$ leading
tidal stream. The predicted radial velocity for this stream also
agrees very well with that of Pal 12 and M99.
There is no M99 data point in the velocity along Galactic longitude and
latitude, since there is no proper-motion determination for this
sample of stars. The velocity of Pal 12 along latitude 
falls within the
predicted velocity of the same $(n_{p}=-2)$ leading stream, and so does the
total energy of Pal 12.  Here the amount of energy difference with respect
to Sgr, is $\sim 16\%$, a value that is slightly smaller than that
reported in Sections 6.3 and 7.1. This is due to the fact  that the present 
calculation was done using the J99 code which has
the JSH95 potential model implemented rather than our preferred P90 model.
One discrepancy we find between the $(n_{p}=-2)$ leading stream and Pal 12
is in the velocity along Galactic
longitude: Pal 12 seems to move faster than the stream.
There are a few possible sources of this discrepancy. The analytical model 
that describes the location and the kinematics of the tidal streams
(Johnston 1998) approximates that the azimuthal period is a function 
only of the orbital energy, and thus directly relates the orbital-energy
change to the azimuthal period for the debris. Along with this assumption, the 
azimuthal period is derived 
in a purely logarithmic potential rather than the 
three-component one. These assumptions work best for satellites 
on circular orbits at relatively large Galactocentric radii ($\sim 30$ kpc), 
where the halo potential dominates. Since Sgr is on a relatively low-energy,
short-period orbit ($R_p = 14$ kpc, Table 4) 
some of these approximations may not
accurately describe the kinematics. Another possible source is the 
Galactic potential model: the tidal stream analysis uses the JSH95
model,  while in Section 7.2
we showed that the preferred model is the P90 one, which provides a 
slightly different orbital-energy scale and a different precession rate.
Another source of adjustment for the precession rate 
is the degree of flatness of the halo potential, which was assumed
spherical in both models used in this paper.
And lastly, the proper motion of Sgr, which is
poorly constrained in longitude at least for Sgr2, 
may also be a source for the discrepancy in the velocity along longitude.

Interestingly, from the tidal stream analysis, Pal 12 appears to have
been torn from
Sgr at $n_p=-2$ which corresponds to $t = -1.4$ Gyr (the radial period is
$P_r = 0.7$ Gyr). From the analysis in Section 7.2 we obtain,
that this event occurred at $t = -1.7$ Gyr in model P90
($P_{r,P90} = 0.9$ Gyr, see Table 4), and
at $t = -1.4$ Gyr in JSH95 model ($P_{r,JSH95} = 0.7$ Gyr).

\subsection{Pal 12 as a Sgr Cluster}

Based on radial velocities and distances with respect to the Sun,
Da Costa \& Armandroff (1995) argued that clusters M54 (NGC 6715),
Ter 7, Arp 2 and Ter 8 are associated with Sgr.
In Table 5 we summarize the metallicities, absolute integrated magnitudes and
concentration parameters for these four clusters and for Pal 12. The data 
are taken from Harris (1996) catalog. Pal 12's concentration parameter
is that derived  by R98. Apart from M54, which is 
significantly more massive and has often been suggested to represent 
the nucleus of Sgr (e.g., Da Costa \& Armandroff 1995, 
Layden \& Sarajedini 2000),
the other four clusters are low-mass, low-concentration clusters.
Pal 12's  metallicity lies in the gap between the metal poor
clusters Ter 8 and Arp 2 and the metal rich cluster Ter 7
while Pal 12's mass is slightly smaller than that of the 
three low-mass clusters (see absolute magnitudes in Table 5). 
Pal 12's  concentration parameter is well within
the range defined by the three low-mass Sgr clusters.
With an age of a few Gyr (R98), Pal 12
falls on the age-metallicity sequence defined by clusters and stars of 
Sgr (see Fig 19 in Layden and Sarajedini 2000).
Thus Pal 12's properties fit nicely within the Sgr cluster 
population.

\section{Discussion}

We have  presented our measured tangential velocity for Pal 12.
Based on proper motion alone, Pal 12's motion could be consistent with 
association with the LMC (as suggested by LR92) or Sgr. However, a more
in depth dynamical analysis reveals a greater likelihood of an 
association with Sgr. 
We have explored in a number of ways the possibility that
Pal 12 was once part of the Sgr system. Our summarized arguments
for this scenario are:

1) Pal 12 is located at the same Galactocentric radius as Sgr,
relatively close in Galactic longitude to Sgr, and towards the
southern extension of Sgr. At this location, Pal 12 is moving towards
Sgr.

2) The comparison of Pal 12 and Sgr orbits shows that,
for our preferred potential model and Sgr proper motion,
there is a minimum in the phase-space distance of the two objects
at $t \sim 1.7$ Gyr ago. Given the uncertainties in the orbits
due to measurement uncertainties, and the fact that Pal 12's
orbit is less energetic than Sgr's orbit at a $\sim 20\%$ level, 
we show that this minimum can satisfy the criteria for a 
bound Pal 12-Sgr system. For the potential model that was used
in the tidal disruption analysis (Section 7.3), we also obtain
a less significant minimum, at  $t \sim 1.4$ Gyr ago, because the
periods are slightly different between the two potential models.

3) From a semi-analytical model of the tidal disruption 
of Sgr (Johnston 1998, J99) we show that Pal 12's 
location, two components of its velocity vector, and
its total orbital energy match those predicted for a
leading tidal stream torn from Sgr two pericentric passages ago,
or $\sim 1.4$ Gyr ago. The timing of this event coincides with that
obtained from the direct comparison of the orbits. One
disagreement we obtain between the model prediction for the
stream and Pal 12 is in the velocity component along 
Galactic longitude.
Possible sources of this disagreement may be in some of
the approximations of the disruption model,
the details of the Galactic potential model, and/or
the poorly constrained motion of Sgr in longitude.

4) Properties of Pal 12 such as age, metallicity, mass, and
concentration parameter fall within those of the three low-mass
clusters associated with Sgr; Ter 7, Ter 8 and Arp 2.

We have also discussed a space-based determination of the Sgr 
proper motion (Sgr1). If one chooses this value over
the less-precise, ground-based one, then it is less likely that
Pal 12 was torn from Sgr. As this Sgr1 proper-motion determination is
the first space-based measurement of an absolute proper motion we
feel more work, ground based and/or space based, is needed to provide a
definitive value.
Starcount surveys mapping Sgr across the area between Pal 12 and 
Sgr's presently known extent are also desirable in order 
to better understand the structure
of its possible debris.

We thank Allan Sandage and the Observatories of the Carnegie Institution 
for loan of the Hale 5-m plates taken by himself and Baade. We also thank 
Peter Stetson for loan of the CTIO plates. We are grateful to
 Kathryn Johnston for lending the code for the tidal disruption of Sgr,
as well as for many useful suggestions and comments. 
This research was supported by NSF grant AST-9702521.

\newpage

\begin{figure} 
\caption{Proper-motion errors in one coordinate as a function of
magnitude (top panel), and proper motions in $x$ and $y$ as a function of
magnitude (middle and bottom panel respectively). Filled circles
represent red giants which are clusters members based on radial velocities.
The filled triangles represent the blue stragglers identified in R98.}
\end{figure}

\begin{figure}
\caption{Proper motion of galaxies versus positions, magnitude and
color. The top panels, that have crosses as symbols, show
the proper motions as derived from a global plate solution. The remaining
panels show the proper motions as derived from a local solution.}
\end{figure}

\begin{figure}
\caption{Comparison between the orbit of Pal 12 and that of Sgr.
The first (top) panel shows the difference in position between the
two objects as a function of time, the second panel shows the difference
in velocity, the third panel shows the phase-space normalized ``distance''
and the fourth (bottom) panel shows the $\Delta r \Delta V^2$ quantity.
The thick lines correspond to the Sgr2 proper motion (Table 3), while the
thin ones to the Sgr1. The continuous lines represent
the integration in the potential model P90,
while the dashed lines that in the potential model JSH95.}
\end{figure}

\begin{figure}
\caption{Simulations of orbit-combinations as generated from
Gaussian errors in the observables. The left panel shows 
$\Delta E_{orb}$ as a function of $d_{min}$ as
defined by equation 1. The right 
panel shows $d_{min}$ as a function of time for the restricted range
in the orbital-energy change: from -22 \% to -18\%.}

\end{figure}

\begin{figure}
\caption{Tidal streams from Sgr: positions and kinematics at the present
time. The two top panels show the distribution in the Galactic plane and
perpendicular to the Galactic plane. In all panels the filled triangle 
represents Sgr, the filled circle Pal 12, and the filled square the 
candidate Sgr red clump stars from M99. The leading streams are
represented with small filled circles, and the trailing ones with 
open circles. The darkest shade represents the most recent pericentric
passage, while lighter shades represent correspondingly previous 
pericentric passages. The two top panels show the spatial distribution
of the streams and our three objects. The continuous line is the orbit
of Sgr, and the dashed line that of Pal 12. The following panels show the
velocity components and orbital-energy difference as a function of 
Galactic latitude.}
\end{figure}
\newpage

{\sc TABLE} 1. Plate Material

{\sc TABLE} 2. Refraction and Distortion Corrections

{\sc TABLE} 3. Absolute Proper Motions

{\sc TABLE} 4. Orbital Parameters

{\sc TABLE} 5. Cluster Parameters

\end{document}